# Chemical manipulation of hydrogen induced high p-type and n-type conductivity in Ga$_2$O$_3$

Md Minhazul Islam[1], Maciej Oskar Liedke[2], David Winarski[1], Maik Butterling[2], Andreas Wagner[2], Peter Hosemann[3], Yongqiang Wang[4], Blas Uberuaga[4] and Farida A. Selim[1] *

[1]Center for Photochemical Sciences, Bowling Green State University, Bowling Green, Ohio 43403, USA.

[2]Institute of Radiation Physics, Helmholtz-Center Dresden-Rossendorf, Dresden 01328, Germany.

[3]Department of Nuclear Engineering, University of California at Berkeley, Berkeley, CA 94720, USA.

[4]Materials Science and Technology Division, Los Alamos National Laboratory, Los Alamos, New Mexico 87545, USA.

**ABSTRACT:** Advancement of optoelectronic and high-power devices is tied to the development of wide band gap materials with excellent transport properties. However, bipolar doping (n-type and p-type doping) and realizing high carrier density while maintaining good mobility have been big challenges in wide band gap materials. Here P-type and n-type conductivity was introduced in β-Ga$_2$O$_3$, an ultra-wide band gap oxide, by controlling hydrogen incorporation in the lattice without further doping. Hydrogen induced a 9-order of magnitude increase of n-type conductivity with donor ionization energy of 20 meV and resistivity of $10^{-4}$ Ω.cm. The conductivity was switched to p-type with acceptor ionization energy of 42 meV by altering hydrogen incorporation in the lattice. Density functional theory calculations were used to examine hydrogen location in the Ga$_2$O$_3$ lattice and identified a new donor type as the source of this remarkable n-type conductivity. Positron annihilation spectroscopy confirmed this finding and the interpretation of the results. This work illustrates a new approach that allows a tunable and reversible way of modifying the conductivity of semiconductors and it is expected to have profound implications on semiconductor field. At the same time it demonstrates for the first time p-type and remarkable n-type conductivity in Ga$_2$O$_3$ which should usher in the development of Ga$_2$O$_3$ devices and advance optoelectronics and high-power devices.

## I. INTRODUCTION:

A wide band gap energy has become a key parameter for the future development of high-power transistors and optoelectronic devices;[1,2] and wide band gap oxides, such as ZnO, have been shown to exhibit excellent characteristics. However, their deployment in many applications has been hindered due to the lack of conductivity control or the difficulty of realizing high carrier density with good mobility. Bipolar doping (realizing both n-type and p-type) is one of the big challenges in wide band gap materials but it is crucial for most devices.[3] Further, substitutional doping of elements, the common method to provide charge carriers, often causes disorder, suppressing carrier mobility and there is always a trade-off between increasing the maximum attainable carrier density and maintaining good mobility in oxides. In this work, we report how to induce p-type and n-type conductivity in an ultra-wide band gap oxide (Ga$_2$O$_3$) through controlling hydrogen (H) incorporation in the lattice without further substitutional doping and demonstrate a charge carrier density of $10^{21}$ cm$^{-3}$ with electron mobility 100 cm$^2$ V$^{-1}$ S$^{-1}$ at room temperature leading to $10^{-4}$ Ω.cm resistivity. Such high electron density and good mobility is remarkable for oxide semiconductors. We identify a new donor concept behind this remarkable conductivity. For p-type conductivity we report a hole density of $10^{20}$ cm$^{-3}$, but with very low hole mobility less than 1, which is expected from the flat valence band in Ga$_2$O$_3$.

The study was carried out on Ga$_2$O$_3$, which is emerging as a promising material for high power devices due to its large band gap (~ 4.5-5eV) and high breakdown field of 8 MV/cm; it is receiving significant attention in the scientific community as a potential candidate for a wide range of applications.[4-9] β-Ga$_2$O$_3$ is the most stable polymorph of the Ga$_2$O$_3$ phases, with a monoclinic crystal structure of space group $C2/m$.[7] It behaves as an insulator in its defect free crystalline form. As of today, only one type of conductivity (n-type) has been achieved by doping β-Ga$_2$O$_3$ with Sn, Ge or Si during growth.[10-13] With respect to p-type conductivity, there has not been any significant success.

Hydrogen is known to have a strong influence on the electrical conductivity of semiconductors.[14] In β-Ga$_2$O$_3$ monoatomic H has a low formation energy and can occupy both interstitial and substitutional sites to act as a shallow donor.[15] The complex crystal structure of β-Ga$_2$O$_3$ allows for the formation of many configurations where interstitial hydrogen (H$_i^+$) forms a bond with oxygen, creating electronic states which are close in energy. According to J. Varley et.al.[9], H$_i$ acts as a shallow donor and substitutional hydrogen, H$_O$, has low formation energy only under oxygen poor condition. Despite these theoretical predictions on the possibility of n-type conductivity due to H-incorporation in various locations, there has not been any report on significant experimental success. In this work, we generate H-donors and H-acceptors in Ga$_2$O$_3$ by controlling H incorporation on cation vacancy sites, not as H$_i$ or H$_O$. A cation vacancy is an electrical compensating acceptor in semiconductors including β-Ga$_2$O$_3$.[16] Although cation vacancies have high formation energy in some oxide semiconductors (e.g. SnO$_2$, In$_2$O$_3$), previous first principle calculations showed that their formation energy is significantly lower in β-Ga$_2$O$_3$ and hence a high probability of H-decorated V$_{Ga}$ formation can be achieved after incorporating H into the crystal.[16,17]

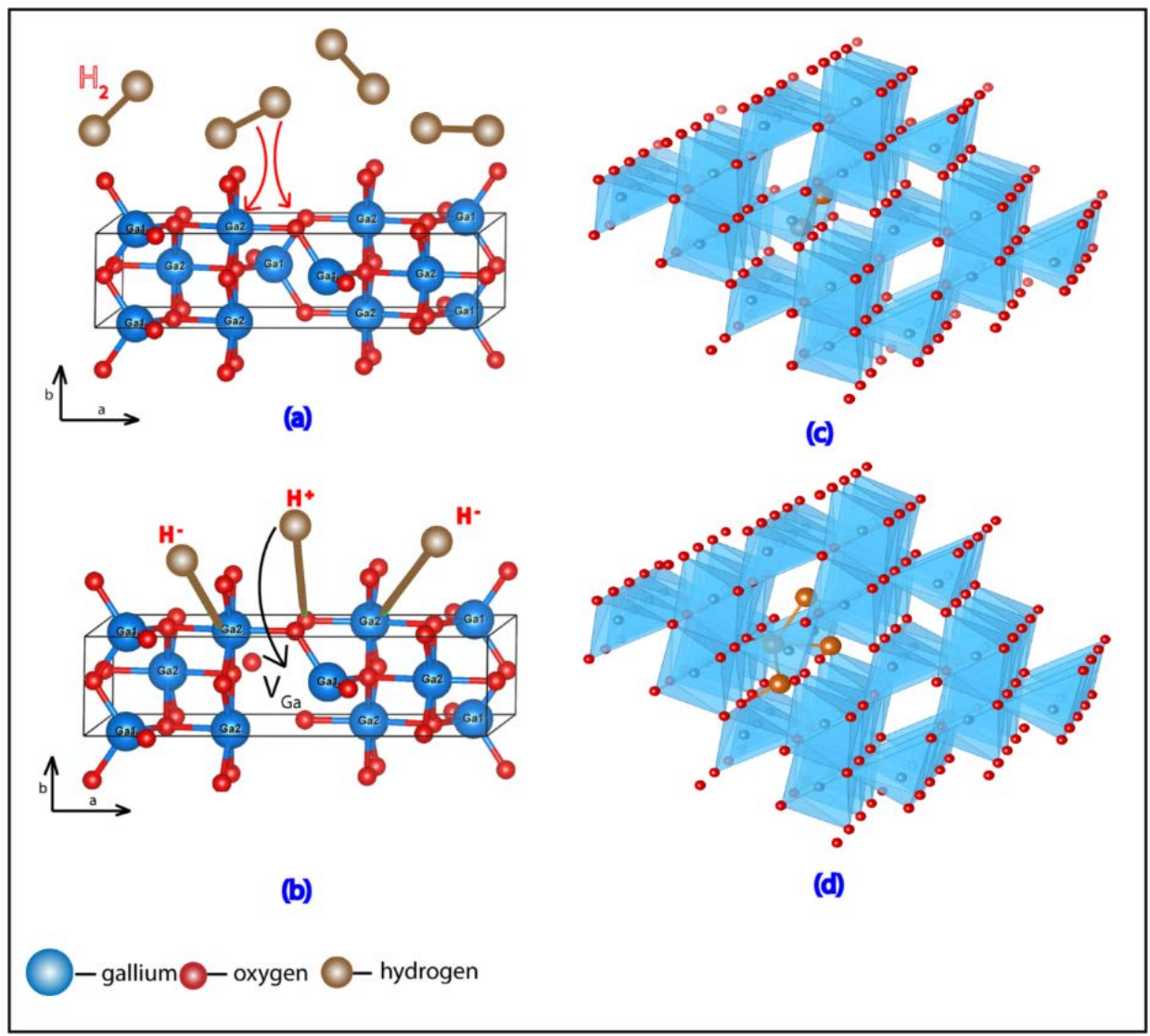

Figure 1: Schematic diagrams showing hydrogen incorporation in β-Ga$_2$O$_3$ (a) hydrogen molecules coming in contact with the surface at elevated temperature and dissociating heterolytically. The electron cloud of H$_2$ is attracted toward gallium while the proton is attracted toward oxygen. (b) The proton and hydride ion are attached to oxygen and gallium atoms, respectively, on the crystal surface and diffuse through the bulk crystal at high temperatures. The proton is attracted toward the negatively charged gallium vacancy. (c) Ga vacancy decorated with two hydrogen as predicted from DFT calculations providing stable acceptor state (d) Ga vacancy decorated with four hydrogen as predicted from DFT calculations providing stable donor state.

It is necessary to understand the interaction of H$_2$ with the surface of metal-oxide semiconductors to gain insight on the process of H-incorporation into the crystal. H-incorporation into the crystals at high temperature occurs in two steps. At first, H$_2$ dissociates and becomes attached to the surface, then diffuses into the bulk crystal. Depending on the nature of the materials, H$_2$ can follow either homolytic or heterolytic dissociation pathways. In case of homolytic cleavage, H$_2$ molecule dissociates to form two H-atoms that become attached to the oxygen on the crystal surface. On the other hand, H$_2$ dissociates to form a proton and a hydride during heterolytic cleavage where the proton and hydride become attached to the oxygen and metal atoms respectively. The redox capacity of metals determines the type of dissociation that is most likely to occur. Density functional theory (DFT) predicts that H$_2$ tends to dissociate heterolytically on nonreducible oxide (e.g MgO, γ-Al$_2$O$_3$) surfaces while following a homolytic pathway on reducible oxide (e.g. CeO$_2$) surfaces.[18] β-Ga$_2$O$_3$ was found to be nonreducible via DFT.[19] Therefore, it is most likely that H$_2$ follows heterolytic dissociation as shown in Fig. 1a. The adsorbed proton and hydride diffuse into the bulk crystal at high temperatures. The proton is attracted toward the negatively charged V$_{Ga}$ while the hydride is attracted toward the positively charged or neutral V$_O$, as shown in Figure 1b.

**Table 1.** Transport properties of Ga$_2$O$_3$ samples, the thickness of the conductive layer for p-type and n-type is 500 nm

| (a) H$_2$ diffusion took place in a closed ampoule at 700°C and 580 torr for one hour | | | |
|---|---|---|---|
| sample number | sample | sheet number (cm$^{-2}$) | sheet resistance (ohm/cm$^2$) |
| 1 | undoped β-Ga$_2$O$_3$ single crystal | 7.00E+06 | 1.940E+8 |
| 2 | annealed in H$_2$ | 5.45E+10 (P- type) | 1.480E+5 |
| 3 | annealed in H$_2$ (after 4 days) | 3.44E+06 | 7.330E+8 |
| 4 | annealed in H$_2$ (2nd time) | 1.54E+15 (P-type) | 4.060E+1 |
| 5 | annealed in H$_2$ (2nd time, after 4days) | 3.24E+06 | 2.360E+8 |
| (b) H$_2$ diffusion took place in a closed ampoule at 950°C and 580 torr for two hours | | | |
| sample number | sample | sheet number (cm$^{-2}$) | sheet resistance (ohm/cm$^2$) |
| a | undoped β-Ga$_2$O$_3$ single crystal | 5.67E+06 | 3.151E+7 |
| b | annealed in H$_2$ (immediately after annealing) | 1.20E+15 (p-type) | 1.288E+1 |
| c | annealed in H$_2$ (4 days after annealing) | 1.35E+15 (p-type) | 4.126E+1 |
| (c) samples annealed in different environments at 950°C for two hours followed by H$_2$ diffusion at the same temperature and pressure (580 torr) | | | |
| sample number | sample | sheet number (cm$^{-2}$) | sheet resistance (ohm/cm$^2$) |
| 1 or a | as-grown undoped β-Ga2O3 single crystal | 5.67E+06 | 3.15E+7 |
| 2 | annealed in O$_2$ | 2.87E+06 | 1.99E+9 |
| 3 | annealed in O$_2$ followed by annealed in H$_2$ | 6.14E+16 (n-type) | 6.21E+0 |
| b | annealed in Ga followed by annealed in H$_2$ | 1.55E+10 | 2.59E+5 |

## II. MATERIALS AND METHODS:

### A. HYDROGEN INCORPORATION PROCESS:

High quality β-Ga$_2$O$_3$ samples grown by Edge- defined Film-fed Growth (EFG) method were obtained from Tamura Inc., Japan. A number of samples (5mm×5mm×0.5mm) were placed in a quartz ampoule with one open end that was connected to a vacuum pump to pump the air out and evacuate the ampule. After that, the tube was filled with H$_2$ gas at 580 torr pressure. After filling the tube with hydrogen, the open end was properly sealed. The ampoule was placed in an oven where temperature can be precisely controlled. The temperature was increased in two steps up to the desired value and H$_2$ was allowed to diffuse into the crystal for 1 or 2 hours. A few other samples of same dimensions were first annealed in oxygen flow at 950°C and then hydrogen following the same procedure, while others were annealed first with gallium, then hydrogen following the same procedure.

### B. HALL-EFFECT MEASUREMENTS:

Van der Pauw Hall-effect measurements were performed to determine the electrical transport properties of the samples. The measurements were carried out from 30 K to room temperature (298 K) and at constant magnetic field of 9300 G. Four indium contacts were made in a square arrangement on the surface of each sample and carefully adjusted to keep the contacts as small as possible. Current-voltage linearity was checked every time to make sure that the contacts were good and resistivity does not vary more than 10% between different contact points. Temperature dependent measurements of the carrier concentration were carried out from 30 K or below to room temperature using a closed cycle cryostat.

### C. COMPUTATIONAL ANALYSIS:

Density functional theory, as implemented in the Vienna ab-initio Simulation Package (VASP),[20,21] was used to examine H-

incorporation into a Ga-vacancy. These calculations were performed on a 1x4x2 supercell of β-Ga$_2$O$_3$, containing a total of 160 atoms in the defect-free structure. A Γ-centered 2x2x2 Monkhorst-Pack k-point mesh[22] was used to sample the Brillouin zone. The energy cutoff for the planewaves was 400 eV. Pseudopotentials based on the projector augmented wave method[23] and the Perdew, Burke, and Ernzerhof (PBE)[24] generalized gradient approximation (GGA) exchange-correlation functional were used. Calculations were continued until the maximum component of the force on any atom was less than 0.02 eV/angstrom, with one exception (the charged Ga-vacancy), where such a tight convergence was not possible. In this case, the maximum force was 0.024 eV/angstrom. Both monopole corrections (using a calculated dielectric constant of 4.16, which is a bit higher but similar to previously reported values)[25] and an alignment correction were applied to the energies. Instead of averaging the potential to perform the alignment correction, we simply shifted the density of states such that the deepest state in the material aligned across different structures, which has been shown to give similar corrections.[26] In any case, the magnitude of this correction was no greater than 0.1 eV.

A $V_{Ga}$ was created by removing a tetrahedrally-coordinated Ga ion from the cell, as this vacancy structure has been identified as being more favorable.[27] A net charge of -3 was imposed on the structure. H$^+$ ions with a charge of +1 were inserted into the resulting vacancy structure (leaving the total number of electrons in the system constant but reducing the net charge of the cell). The resulting binding energy for each configuration was computed via the following relationship:

$$E_b = E([V_{Ga}NH]^{(-3+N)}) + E(\text{Bulk Ga}_2\text{O}_3) \, E(V_{Ga}^{3-}) - NE(H^+) \quad (1)$$

where $E([V_{Ga}NH]^{(-3+N)})$ is the energy of the system with the Ga vacancy filled with $N$ H$^+$ ions, $E(\text{Bulk Ga}_2\text{O}_3)$ is the energy of defect-free β-Ga$_2$O$_3$, $E(V_{Ga}^{3-})$ is the energy of the isolated Ga vacancy in a 3- charge state, and $E(H^+)$ is the energy of an isolated 1+ H interstitial. With this definition, a negative energy indicates an exothermic or favorable reaction. We did not perform a systematic search for the lowest energy H interstitial position, but performed multiple minimizations where the H was randomly displaced to find a reasonable structure. The structure found here, in which the H$^+$ ion is bonded to one of the three-fold coordinated oxygen ions, is similar to that described by Varley et al.[9]

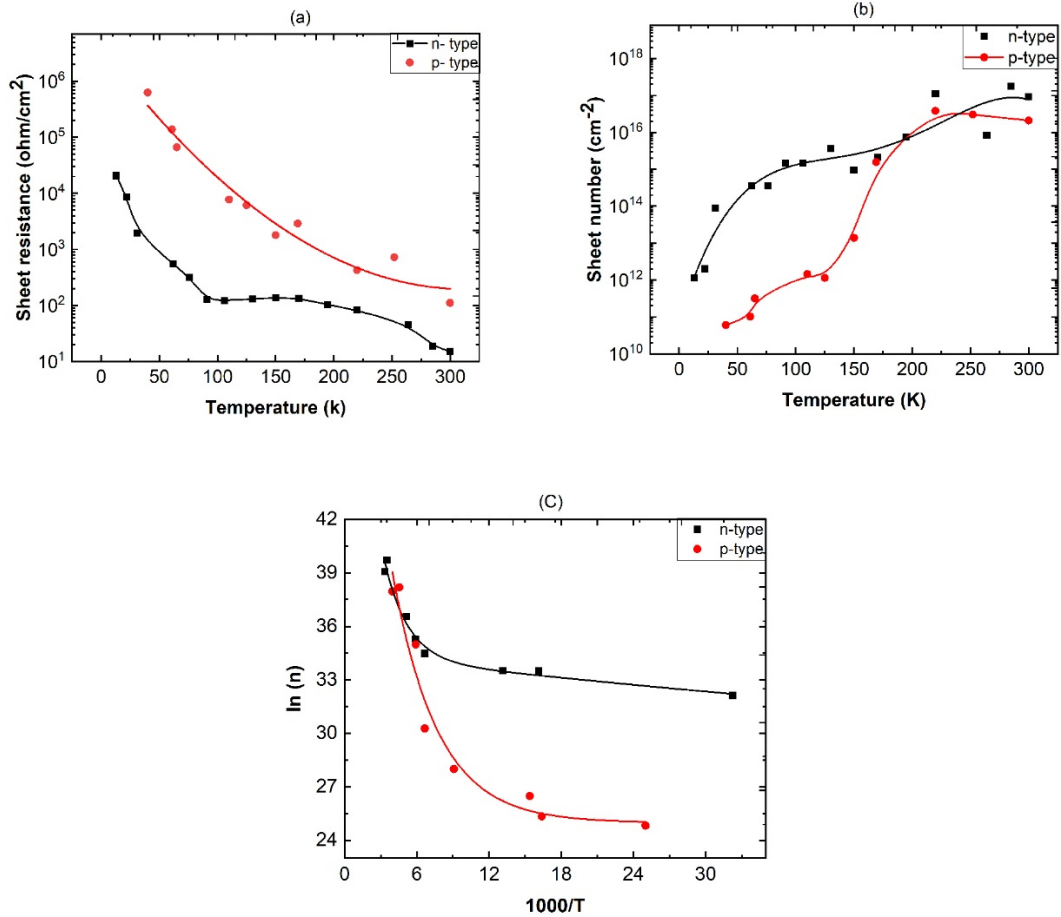

Figure 2: Temperature dependent transport properties of the n-type and p-type H$_2$ treated Ga$_2$O$_3$ samples. (a) sheet resistance, (b) sheet number, (c) sheet number logarithm plotted as a function of 1000/T

### D. THERMAL STIMULATED LUMINESCENCE SPECTROSCOPY (TSL):

Thermal stimulated luminescence (TSL) spectroscopy[28-32] was performed on the samples to calculate the donor and acceptor ionization energies[30]. The measurements were performed using an in-house built spectrometer,[28,33] from -190°C to 25°C. The samples were first placed in a dark compartment and irradiated with UV light at -190°C for 30 min. After irradiation, the temperature of the samples was set to increase at constant rate (60°C/min) and the emission spectra were recorded from 200 to 800 nm at every 5 seconds. The glow curves which represent the emission intensity as a function of temperature were constructed from the integration of emission over wavelengths at each temperature.

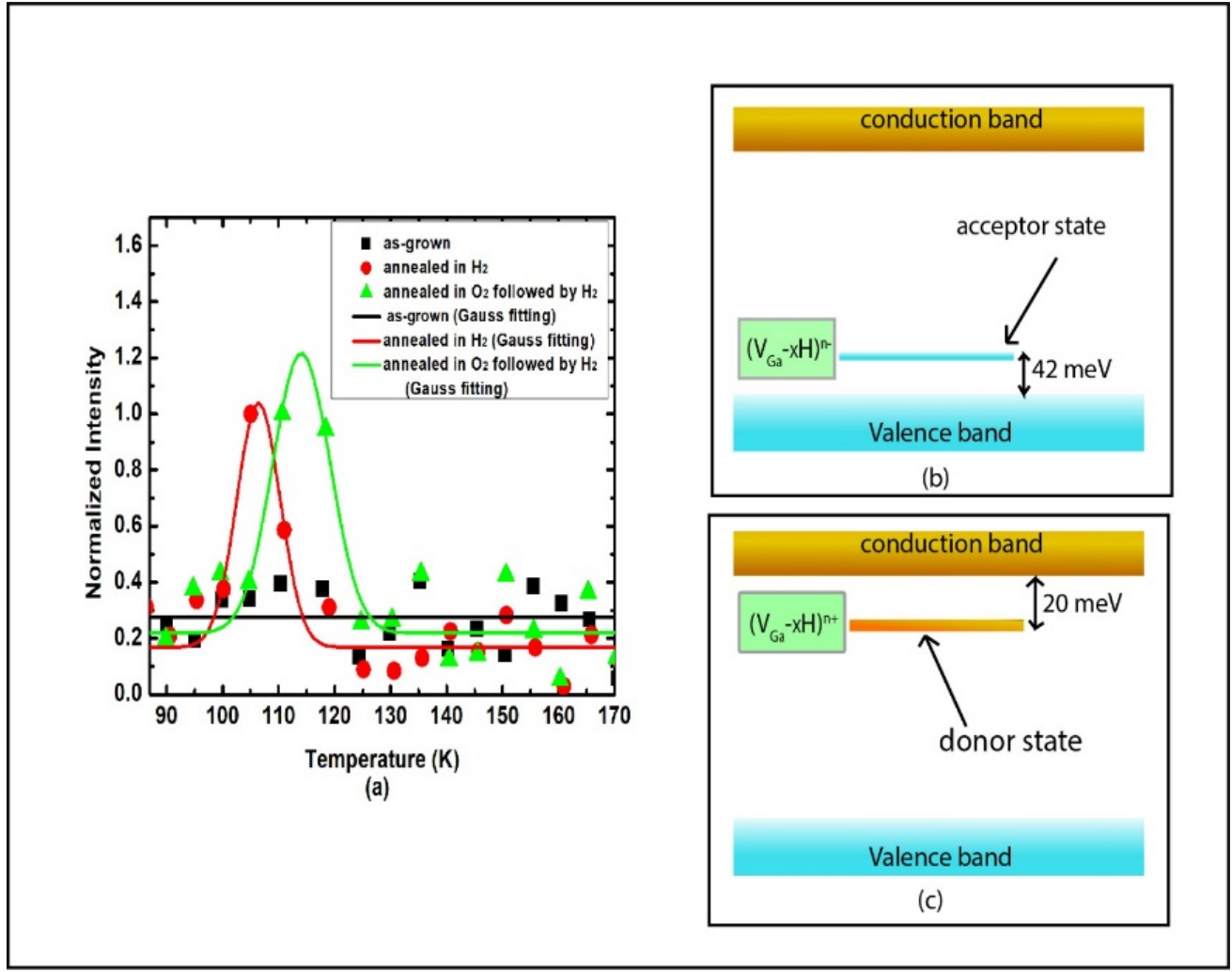

Figure 3: Thermally stimulated luminescence emission (a) of the samples annealed at 950°C for two hours in different environments. Data points for annealed samples were normalized from 0 to 1. Data points for as-grown sample were normalized from 0 to 0.5 to minimize noise (no glow peak). Peaks were fitted with a Gaussian function. The two peaks appeared at low temperature after $H_2$ diffusion, and after O-anneal followed by H-diffusion are associated with the induced shallow acceptor and shallow donor in the samples respectively and they were used for calculating the ionization energies. The flat band diagrams showing donor and acceptor states of the samples after direct hydrogen diffusion (b) and hydrogen diffusion after filling up oxygen vacancies (c).

### E. POSITRON ANNIHILATION SPECTROSCOPY:

We carried out positron annihilation spectroscopy (PAS), which is a well-established technique to detect and characterize cation vacancies in semiconductors.[34-36] Both Doppler Broadening of Positron Annihilation Spectroscopy (DBPAS) and Positron Annihilation Lifetime Spectroscopy (PALS) were employed. DBPAS measurements were carried out using a monoenergetic variable energy positron beam at HZDR facility in Dresden, Germany.[37] Positrons are emitted from an intense $^{22}$Na source and a tungsten moderator and accelerated to discrete energy values $E_p$ in the range of $E_p$=0.05-35 keV. Such positron implantation energy, $E_p$ allows penetrating up to about 1.8 μm in $Ga_2O_3$. Doppler broadened spectra representing positron annihilation distribution for each $E_p$ were acquired using a single high-purity germanium detector with energy resolution of 1.09 ± 0.01 keV at 511 keV and the S and W parameters (defined in the caption of Fig.4) were calculated from the peak. PALS has been established as the most effective method to probe cation vacancy related defects, distinguishing between their types and providing information about their concentrations.[34] PALS was performed at the Mono-energetic Positron Spectroscopy (MePS) pulsed beam, which is the end station of the radiation source ELBE (Electron Linac for beams with high Brilliance and low Emittance) at HZDR facility in Dresden Germany.[37] The lifetime spectrum was measured at each positron energy $E_p$ up to 16 keV with a time resolution of 205 ps. All lifetime spectra contained at least $5\times10^6$ counts and were analyzed as a sum of time-dependent exponential decays, $N(t)=\Sigma_i\ I_i/\tau_i\cdot exp(-t/\tau_i)$ convoluted with the Gaussian's functions describing the spectrometer timing resolution, using the PALSfit fitting software.[38] Depth-resolved measurements of PALS revealed two major positron lifetime components for each sample.

### III. RESULTS AND DISCUSSIONS:

As-grown samples were highly resistive, but after $H_2$-diffusion they showed an increase in carrier density and p-type conductivity. $H_2$-diffusion at 700 °C for 1 hr led to unstable conductivity that decays with time (Table I a). However, $H_2$-diffusion at 950 °C for 2 hrs led to a greater increase in carrier density and stable p-type conductivity over time (Table I b). Other procedures were carried out to incorporate $H_2$ into different sites in the undoped β-$Ga_2O_3$. One sample was annealed in $O_2$ flow and another was annealed with Ga in a closed ampoule at 950 °C for 2 hrs. This process should fill up the respective (anion or cation) vacancies. After that, hydrogen was diffused into the crystals at 580 torr in a closed ampoule at 950 °C for 2 hrs. $O_2$–annealing followed by $H_2$ diffusion led to high n-type conductivity (stable over time) and remarkable sheet carrier density of about $10^{16}$ cm$^{-2}$ with electron mobility 100 cm$^2$/Vs (Table I c). The thickness of the conductive layer where H diffusers in is 500 nm as revealed from positron measurements in Fig. 4a, showing remarkable conductivity ($10^{-4}$ Ω.cm). The sample exhibits 9-orders of magnitude increase in conductivity and 10-orders magnitude increase in carrier density. In contrast, annealing in Ga followed by $H_2$ diffusion did not lead to a significant increase in conductivity (Table I c). Both sole H-diffusion and H-diffusion after $O_2$-anneal treatments were carried on other as-grown undoped $Ga_2O_3$ samples and led to the same results. Samples preserved their p-type or n-type conductivity with no decay or negligible decay after months. Figure 2 shows the temperature dependence of sheet resistance and sheet number of the p-type and n-type $Ga_2O_3$ samples, signifying the ionization of carriers region followed by extrinsic semiconductor behavior at higher temperatures. Intrinsic semiconductor behavior cannot occur at room temperature as band to band transitions are not possible in $Ga_2O_3$ at this temperature because of the ultra-wide band gap.

The realization of p-type and n-type conductivity after $H_2$ diffusion can be explained as follows. A Ga-vacancy acts as a deep acceptor with -3 charge state $(V_{Ga})^{3-}$. During the diffusion of hydrogen into the crystal, the surface adsorbed proton (re Fig 1a-b) becomes attracted toward the $(V_{Ga})^{3-}$ where it stabilizes the negative charge and, therefore, lowers the acceptor state. This results in H-decorated Ga-vacancy $(V_{Ga}-2H)^{1-}$ (as represented in Fig. 1c) and p-type conductivity. At lower temperatures (e.g 700$^0$C), protons are less likely to diffuse deep inside the bulk crystal. This results in a decrease in conductivity over time due to the reverse diffusion at room temperature. However, the high p-type conductivity persists over time for the sample exposed to $H_2$ at higher temperature and for a longer period of time due to the diffusion of $H^+$ deeper into the crystal.

The sample that is exposed to the $H_2$ after filling up $V_O$ (after annealing in $O_2$) showed high n-type conductivity. In this case, more H are diffused to the $V_{Ga}$ due to the absence of $V_O$ leading to the formation of $(V_{Ga}-4H)^{1+}$ as represented in Fig. 1d), which acts as a donor. That is, the absence of $V_O$ in this case means that the only available traps for H are $V_{Ga}$, which thus become filled to a greater extent. The contribution of n-type conductivity from $H_i$, is not prominent as filling up $V_{Ga}$ following by H-diffusion shows a negligible increase in carrier concentration. Moreover, it confirms that the H-decorated $V_{Ga}$ are primarily responsible for the induced n-conductivity in the samples. Density functional theory was used to examine H-incorporation into a Ga-vacancy. The results are presented in Table 2. The binding energy of one $H^+$ ion to the Ga-vacancy is -4.4 eV. The DFT calculations reveal that, as N (the number of H ions) increases, at least up to N=4, the reaction remains exothermic, though the strength of the binding, per H atom, decreases. The energy gained by adding the 4$^{th}$ $H^+$ ion is only -0.8 eV, much less than the -4.4 eV gained by adding the 1$^{st}$ $H^+$ ion. If the trend persisted, this suggests that no more than 4 $H^+$ ions can be favorably accommodated into $V_{Ga}$. Thus, these calculations indicate that a single $V_{Ga}$ can accommodate up to 4 $H^+$ ions, changing the net charge of the complex from 3- (when N=0) to 1+ (when N=4), and confirm that $(V_{Ga}-4H)^{1+}$ (Fig. 1d) is more favorable than $H_i^+$. These calculations verified our interpretation of the electrical transport measurements that $(V_{Ga}-4H)^{1+}$ is the dominant donor in the treated highly conductive n-type sample. This cation vacancy filled with the relevant numbers of $H^+$ represents a new type of donor that does not create disorder in the lattice suppressing electron mobility as in the case of standard dopants on substitutional or interstitial sites.

Figure 3a displays the TSL emission for as-grown, p-type and n-type $H_2$ treated $Ga_2O_3$. The as-grown sample shows no peak corresponding to shallow levels. Each of the other two samples shows a peak at low temperature indicating the formation of shallow level. The peak formed at 107 K in the p-type $H_2$-anneal sample (red curve in Fig. 3a) is associated with the formation of shallow acceptors with ionization energy of 42 meV, calculated using the simplified model of TL developed by Randal and Williams.[28,31,34] The ionization energy of the donor, emerging after $O_2$-annealing followed by $H_2$-diffusion (green curve in Fig. 3a), was also calculated by the initial rise method from the peak at 111 K and found to be 20 meV. Figure 3 b and c shows the corresponding flat band diagram and corresponding donor and acceptor state. The details of the calculation of donor/acceptor ionization energy is provided in supporting information.

To further understand the effect of H-incorporation and confirm our interpretation of the origin of conductivity, we carried out PAS measurements. Figure 4a presents S and W (defined in Fig. 4) as a function of depth for the two treated samples. The large values of S at the very beginning of the two curves are common in all PAS measurements, indicating the formation of positronium at the surface. The graph shows a large difference between the two samples in the first 500 nm with lower S values and higher W values for the sample annealed in $O_2$ followed by $H_2$, which exhibits high n-type conductivity. The decrease in S-parameter is an indication for the suppression of positron trapping at cation or neutral vacancies. Thus, these measurements confirm the decrease of negatively charged and neutral vacancies in the $O_2$-annelead followed by $H_2$-diffusion sample. This must be due to filling of Ga-vacancies with more than three H-ions leading to a positive charge state and the formation of a shallow donor as indicated by the immense increase in n-type conductivity. This $(H-V_{Ga})^{1+}$ complex has a positive charge state and cannot trap positrons, leading to the substantial decrease in S-parameter. On the other

Table 2. Binding energy of H⁺ ions to a Ga vacancy

| N | Net charge of the H-V$_{Ga}$ complex | Binding energy (eV) | Binding energy per H (eV) | Binding energy of extra H (eV) |
|---|---|---|---|---|
| 1 | -2 | -4.4 | -4.4 | -4.4 |
| 2 | -1 | -7.5 | -3.7 | -3.1 |
| 3 | 0 | -9.4 | -3.1 | -1.9 |
| 4 | +1 | -10.2 | -2.6 | -0.8 |

Three different values are provided, each providing a different perspective of the interaction. The first, the binding energy, is the value calculated by Eq. 1. The second, the binding energy per H, is the binding energy normalized by the number of H in the complex. Finally, the binding energy of an extra H is the energy difference between the N and N-1 complexes, and represents the energy released by adding the N$^{th}$ H⁺ ion to the complex. The net charge of the complex is also provided.

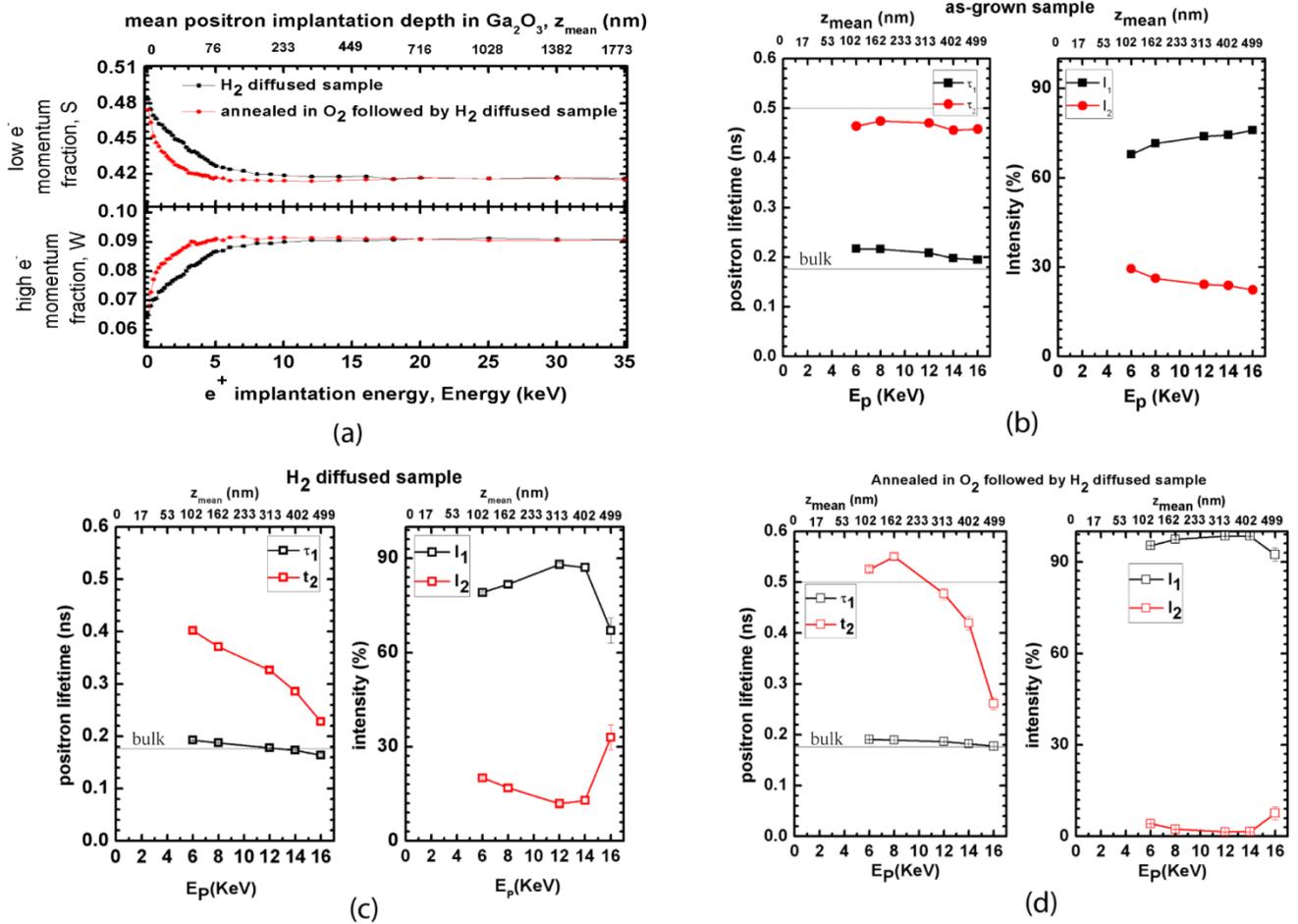

Figure 4: (a) Defect parameters S and W measured by Doppler Broadening of Positron Annihilation Spectroscopy (DBPAS) as a function of penetration depth, S and W are defined as the fraction of positrons annihilating with valence and core electrons respectively. The lower x-axis represents the positron energies and the upper x-axis represents the penetration depth. The graph shows that H$_2$ diffuses about 500 nm in the crystal. Positron Annihilation Lifetime Spectroscopy (PALS) data of (b) as-grown and (c) annealed in H$_2$ (950°C for 2 hours) (d) annealed in O$_2$ followed by H$_2$ (950°C for 2 hours) samples. E$_P$ = Positron implantation energy, Z$_{mean}$= positron implantation depth, τ = positron lifetime, I= intensity of lifetime component, the graphs b, c, and d show the two positron lifetime components and their intensities in each sample.

hand, sole $H_2$-diffusion leads to partial filling of $V_{Ga}$ with hydrogen maintaining a negative charge state and leading to shallow acceptors, which imparts p-type conductivity. This (H-$V_{Ga}$)$^{1-}$ complex is still an active positron trap which leads to a higher S-value.

Depth-resolved measurements of PALS revealed two major positron lifetime components for each sample (supplementary figure 3). Fig. 4b, 4c, and 4d show the lifetime components and their intensity as a function of depth for the as-grown sample, and the $H_2$ diffused, and $O_2$-annealed followed by $H_2$-diffused samples. A distinctive difference can be seen in the intensity and magnitude of the positron lifetime components among the three samples. The large second lifetime component $\tau_2$ indicates the presence of $V_{Ga}$-related defects with negative charge states. For as-grown $Ga_2O_3$, $\tau_2$ is about 470 ps with about 25 to 30% intensity across the sample depth (Fig. 4b). After $H_2$-anneal, $\tau_2$ was reduced to ~320 ps indicating partial filling of $V_{Ga}$ related defects with hydrogen while its intensity was reduced to about 13% (Fig. 4c) due to the decrease of positron trapping at these vacancies as result of less negativity. After annealing in $O_2$ followed by $H_2$-diffusion, almost all positrons annihilate with lifetimes close to the bulk lifetime (Fig. 4d)[39]. The intensity of $\tau_2$ was reduced to about 1% indicating almost complete absence of positron trapping at defects providing strong evidence for filling up $V_{Ga}$ related defects with $H_2$ transforming them into donors with a positive charge state, which cannot trap positrons. Thus, DBPAS and PALS measurements explicitly confirm our interpretation for the origin of n-type and p-type conductivity.

## IV. CONCLUSIONS:

In summary, by controlling H-incorporation in the lattices, we have demonstrated the development of stable p-type and n-type $Ga_2O_3$, which is expected to significantly advance optoelectronics and high-power devices. In the mean time we illustrated a potential simple method for tuning and switching the conductivity of semiconductors between p-type and n-type with the realization of remarkable high carrier density and good mobility in wide band gap oxides, which is a significant challenge by common substitutional doping methods. A concept for new donor type as cation vacancy filled with the relevant numbers of $H^+$ was introduced and found to be behind the remarkable n-type conductivity. This new donor type does not create disorder in the lattice, which often suppresses carrier mobility in the case of standard doping.

## ASSOCIATED CONTENT

**Supporting Information**
Thermoluminescence spectroscopy, Positron Annihilation Lifetime Spectroscopy
This material is available free of charge via the Internet

## AUTHOR INFORMATION

### Corresponding Author

*Corresponding Author: faselim@bgsu.edu

### Notes
The authors declare no competing financial interest.

## ACKNOWLEDGEMENTS
BPU acknowledges helpful discussions with David Andersson and Ghanshyam Pilania. This work was supported as part of FUTURE (Fundamental Understanding of Transport Under Reactor Extremes), an Energy Frontier Research Center funded by the U.S. Department of Energy, Office of Science, Basic Energy Sciences. Los Alamos National Laboratory is operated by Triad National Security, LLC, for the National Nuclear Security Administration of U.S. Department of Energy (Contract No. 89233218CNA000001).

## REFERENCES

[1] Higashiwaki, M., Sasaki, K., Kuramata, A., Masui, T., and Yamakoshi, S. Gallium oxide ($Ga_2O_3$) metal-semiconductor field-effect transistors on single-crystal β-$Ga_2O_3$ (010) substrates. *Appl. Phys. Lett.* 100, 013504 (2012).

[2] Oishi, T., Koga, Y., Harada, K. and Kasu, M. High-mobility β-$Ga_2O_3$(201) single crystals grown by edge-defined film-fed growth method and their Schottky barrier diodes with Ni contact. *Appl. Phys. Express* 8, 031101 (2015).

[3] Ogo, Y., Hiramatsu, H., Nomura, K., Yanagi, H., Kamiya, T., Hirano, M., Hosono, H. p-channel thin-film transistor using p-type oxide semiconductor, SnO. *Appl. Phys. Lett.*, 93, 032113 (2008).

[4] Mastro, M. A., Kuramata, A., Calkins, J., Kim, J., Ren, F., and Pearton, S. J. Perspective-Opportunities and Future Directions for $Ga_2O_3$. *ECS J. Solid State Sci. Technol*. 6, 356 (2017).

[5] Higashiwaki, M. and Jessen, G. H. The dawn of gallium oxide microelectronics. *Appl. Phys. Lett.* 112, 060401 (2018).

[6] Pearton, S. J., Yang, J., Cary, P.H., Ren, F., Kim, J., Tadjer M. J., and Mastro, M. A., Perspective: $Ga_2O_3$ for ultra-high power rectifiers and MOSFETS. *Appl. Phys. Rev.* 5, 011301 (2018).

[7] Geller, S., Crystal Structure of β-$Ga_2O_3$. *J. Chem. Phys.* 33, 676 (1960).

[8] Hajnal, Z., Miro, J., Kiss, G., Reti, F., Deak, P., Herndon, R. C., and Kuperberg, J. M. Role of oxygen vacancy defect states in the *n*-type conduction of β-$Ga_2O_3$. *J. Appl. Phys.* 86, 3792 (1999).

[9] Varley, J., Weber, J., Janotti, A., and Van de Walle, C., Oxygen vacancies and donor impurities in β-$Ga_2O_3$. *Appl. Phys. Lett.* 97, 142106 (2010).

[10] Ueda, N., Hosono, H., Waseda, R., and Kawazoe, H. Synthesis and control of conductivity of ultraviolet transmitting β-$Ga_2O_3$ single crystals. *Appl. Phys. Lett.* 70, 3561 (1997).

[11] Villora, E. G., Shimamura, K., Yoshikawa Y., Ujiie T., and Aoki K. Electrical conductivity and carrier concentration control in β-$Ga_2O_3$ by Si doping. *Appl. Phys. Lett.* 92, 202120 (2008).

[12] Sasaki, K., Higashiwaki, M., Kuramata, A., Masui, T., and Yamakoshi, S. Si-ion implantation doping in β-$Ga_2O_3$ and Its application to fabrication of low-resistance ohmic contacts. *Appl. Phys. Express* 6, 086502 (2013).

[13] Irmscher, K., Galazka, G., Pietsch, M., Uecker, R., and Fornari, R. Electrical properties of *β*-$Ga_2O_3$ single crystals grown by the Czochralski method. *J. of Appl. Phys.* 110, 063720 (2011).

[14] McCluskey, M., Tarun, M., & Teklemichael, S., Hydrogen in oxide semiconductors. *J. of Mat. Res.*, 27, 17 (2012).

[15] Janotti, A. and Van de Walle, C. G. Hydrogen multicentre bonds. *Nature Mater*. 6, 44 (2007).

[16] Korhonen, E., Tuomisto, F., Gogova, D., Wagner, G., Baldini, M., Galazka, Z., Schewski, R., and Albrecht, M. Electrical compensation by Ga vacancies in $Ga_2O_3$ thin films. *Appl. Phys. Lett.* 106, 242103 (2015).

[17] Weiser, P., Stavola, M., Fowler, W., Qin, Y., and Pearton S. Structure and vibrational properties of the dominant O-H center in β-$Ga_2O_3$. *Appl. Phys. Lett.* 112, 232104 (2018).

[18] Garcia-Melchor, M., Lopez, N. Homolytic Products from Heterolytic Paths in H2 Dissociation on Metal Oxides: The Example of $CeO_2$. *J. Phys. Chem. C* 118, 20 (2014).

[19] Helali, Z., Jedidi, A., Syzgantseva, O. A., Calatayud, M., Minot, C. Scaling reducibility of metal oxides. *Theoret. Chem. Acc.: Theory, Computation, and Modeling, Springer Verlag* 136, 9, 100 (2017).

[20] Kresse, G. and Furthmuller, J. Efficient iterative schemes for *ab initio* total-energy calculations using a plane-wave basis set. *Phys. Rev. B* 54, 11169 (1996).

[21] Kresse, G. and Joubert, D. From ultrasoft pseudopotentials to the projector augmented-wave method. *Phys. Rev. B* 59, 1758 (1999).

[22] Monkhorst, H. J. and Pack, J. D. Special points for Brillouin-zone integrations. *Phys. Rev. B* 13, 5188 (1976).

[23] Blöchl, P. E. Projector augmented-wave method. *Phys. Rev. B* 50, 17953 (1994).

[24] Perdew, J. P., Burke, K., and Ernzerhof, M. Generalized Gradient Approximation Made Simple. *Phys. Rev. Lett.* 77, 3865 (1996).

[25] Sturm, C., Furthmüller, J., Bechstedt, F., Schmidt-Grund, R. and Grundmann, M. Dielectric tensor of monoclinic $Ga_2O_3$ single crystals in the spectral range 0.5–8.5 eV *Apl. Mat. 3,* 106106 (2015).

[26] Yadav, S. K., Uberuaga, B. P., Nikl, M., Jiang, C., Stanek, C. R. Band-Gap and Band-Edge Engineering of Multicomponent Garnet Scintillators from First Principles. *Phys. Rev. Appl.* 4, 054012 (2015).

[27] Varley, J. B., Peelaers, H., Janotti, A. and Van de Walle, C.G. Hydrogenated cation vacancies in semiconducting oxides. *J. Phys.: Condens. Matter* 23, 334212 (2011).

[28] Mackay, D. T., Varney, C. R., Buscher, J., and Selim, F. A. Study of exciton dynamics in garnets by low temperature thermo-luminescence. J. Appl. Phys 112, 023522 (2012).

[29] Pagonis, V., Kitis, G., and Furetta, C. Numerical and Practical Exercises in Thermoluminescence. Springer, New York, NY, (2006).

[30] Ji, J., Boatner, L. A., and Selim, F. A. Donor characterization in ZnO by thermally stimulated luminescence. Appl. Phys. Lett. 105, 041102 (2014).

[31] Islam, M. M., Rana, D., Hernandez, A., Haseman, M. and Selim, F. A. Study of trap levels in β-$Ga_2O_3$ by thermoluminescence spectroscopy. J. Appl. Phys. 125, 055701 (2019).

[32] Bos, A. J. J. High sensitivity thermoluminescence dosimetry. Nucl. Instr. Meth. Phys. *B* 184, 3-28 (2001).

[33] Varney, C. R., Mackay, D. T., Pratt, A., Reda, S. M. and Selim, F. A. Energy levels of exciton traps in yttrium aluminum garnet single crystals. J. Appl. Phys. 111, 063505 (2012).

[34] Mahesh, K., Weng, P. and Furetta, C. Thermoluminescence in solids and its applications. Nuclear Technology, Kent, England, (1989).

[35] Krause-Rehberg, R. and Leipner, H. S. Positron Annihilation in Semiconductors. Springer-Verlag (1999).

[36] Schultz, P. J and Lynn, K. G. Interaction of positron beams with surfaces, thin films, and interfaces. Rev. of Mod. Phys. 60 (3), 701, (1988).

[37] A. Wagner, M. Butterling, M. O. Liedke, K. Potzger and R. Krause-Rehberg, AIP Confer. Pro. 1970, 040003 (2018).

[38] Olsen, J. V., Kirkegaard, P., Pedersen, N. J. and Eldrup, M. PALSfit: A new program for the evaluation of positron lifetime spectra. Phys. Status Solidi *C* 4, 4004 (2007).

[39] Ting, W. Y., Kitai, A. H., and Mascher, P., Crystallization phenomena in β-$Ga_2O_3$ investigated by positron annihilation spectroscopy and X-ray diffraction analysis. Mater. Sci. Eng., *B* 91, 541 (2002).

Supplementary Information for

**"Chemical manipulation of hydrogen induced high p-type and n-type conductivity in $Ga_2O_3$"**

## Supplementary Discussion 1:

**Calculations of Donor/Acceptor ionization energy by Thermoluminescence Spectroscopy**

This section provides details about the calculations of donor/acceptor ionization energy by Thermoluminescence Spectroscopy.

Thermoluminescence (TL) is the emission of light from materials upon thermal stimulation after irradiating the sample by ionizing radiation at low temperatures. It is a powerful technique to calculate the energy levels of defects that trap charge carriers (e.g. electrons/hole) at low temperature. The phenomena can be explained by energy band theory of solids.[1] At lower temperatures, most of the charge carriers (e.g. electrons/holes) reside in the valence band in an ideal semiconductor. Electrons can be excited to the conduction band (holes to the valence band) upon excitation. Wide band gap materials often have structural defects that can trap charge carriers. Donor/acceptor states can also be thought of as defects that trap charge carriers at low temperatures. Thermal stimulation can release the electrons/holes from these traps where they transfer their energy to luminescence centers. A schematic diagram of TL process is given in supplementary figure 1 for donor and acceptor cases.

Donor/acceptor ionization energy was calculated by initial rise method.[1] Randall and Wilkins simplified the thermoluminescence model by assuming negligible re-trapping, linear heating rate and formulated the well-known Randall–Wilkins first order expression for TL intensity[1]

$$I(T) = n_0 \frac{s}{\beta} exp\left\{-\frac{E_D}{kT}\right\} \times exp\left\{-\frac{s}{\beta}\int_{T_0}^{T} exp\left\{-\frac{E_D}{kT'}\right\}dT'\right\} \quad (1)$$

Here, s is the frequency factor and is considered as a constant in the simplified model, T is the absolute temperature, k is Boltzman constant and $E_D$ is the donor/acceptor ionization energy, $n_0$ is the total number of trapped electrons/holes at time t=0, β is the constant heating rate. The symmetric shape of the peaks for our samples indicates second or higher order kinetics where significant re-trapping of charge carrier occurs after de-trapping from the traps. A similar equation was derived for the second order kinetics where significant re-trapping occurs.[2]

$$I(T) = \frac{n_0^2 s}{N\beta} exp\left\{-\frac{E_D}{kT}\right\} \times \left[1 + \frac{n_0 s}{N\beta}\int_{T_0}^{T} exp\left\{-\frac{E_D}{kT'}\right\}dT'\right]^{-2} \qquad (2)$$

Initially, intensity of glow peak is dominated by the first exponential half of these equations [ equations (1), (2)] and the last half can be negligible. As a result, if ln(I) is plotted as a function of 1/T for the initial points of the glow peak, a straight line is obtained with the slope from which donor/acceptor ionization energy, $E_D$, can be calculated. Linear fittings of ln(I) vs 1/T for n-type (Fig. 2a) and p-type (Fig. 2b)) samples are shown in Supplementary figure 2. Donor ionization energy of (a) β-$Ga_2O_3$ sample annealed in oxygen followed by hydrogen diffusion and acceptor ionization energy of (b) hydrogen diffused β-$Ga_2O_3$ sample were found to be 20 meV and 42 meV respectively.

REFERENCES

[1]A. J. J. Bos, 'High sensitivity thermoluminescence dosimetry', *Nucl. Instr. Meth. Phys. B* 184, 3-28 (2001).

[2]C. Greskovich and S. Duclos, 'Ceramic scintillators' *Annu. Rev. Mater. Sci*. 27, 69 (1997).

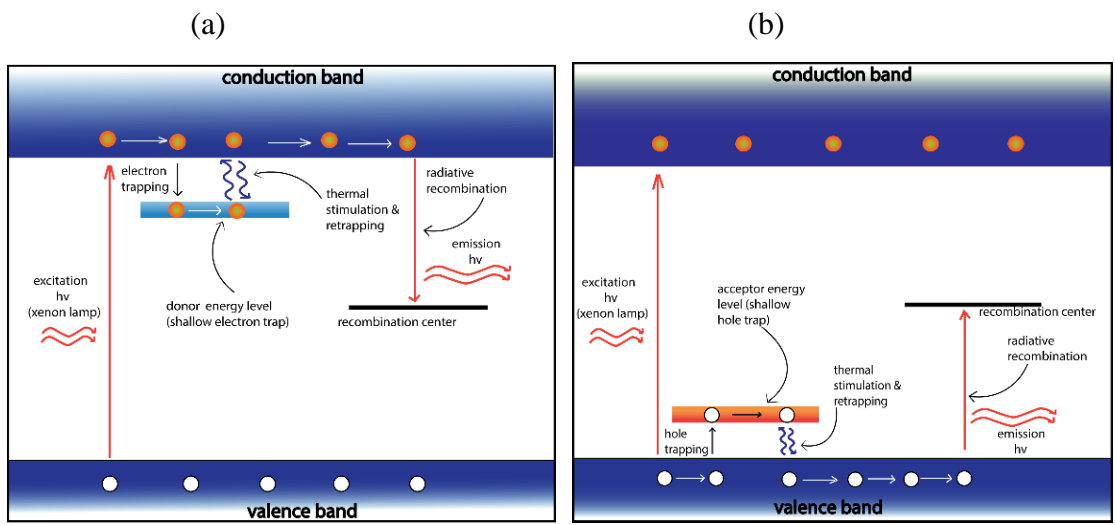

*Supplementary figure 1: Schematic diagram of Thermoluminescence process for (a) donor and (b) acceptor case*

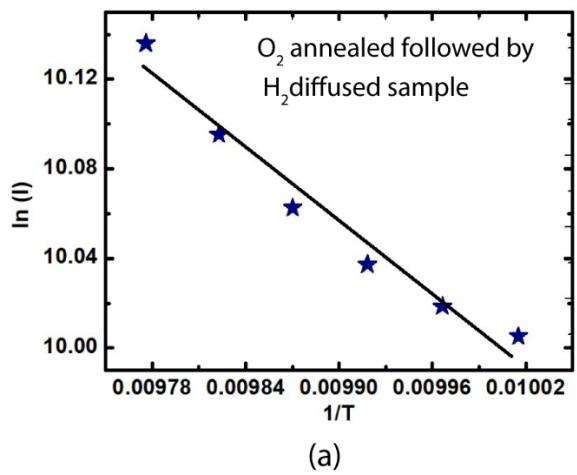 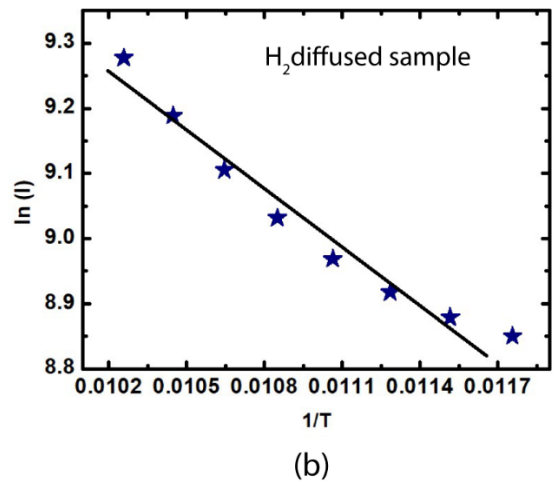

*Supplementary figure 2: Calculation of ionization energy by Initial Rise method. Linear fitting of ln (I) vs 1/T of (a) β-$Ga_2O_3$ sample annealed in oxygen followed by hydrogen diffusion (b) hydrogen diffused β-$Ga_2O_3$ sample.*

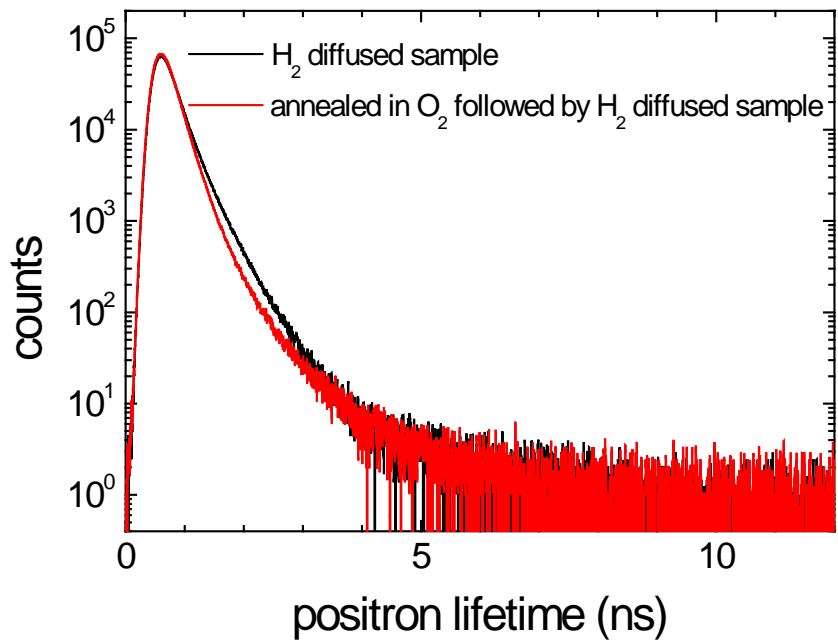

*Supplementary figure 3: Positron lifetime spectra at $E_p$=6 keV for the H$_2$ diffused sample and the sample annealed in O$_2$ followed by H$_2$ diffusion*